\documentclass[
english,			
french,				
spanish,			
brazil,				
]{article}

\usepackage{arxiv}
\usepackage[utf8]{inputenc} 
\usepackage[T1]{fontenc}    
\usepackage{hyperref}       
\usepackage{url}            
\usepackage{booktabs}       
\usepackage{amsfonts}       
\usepackage{nicefrac}       
\usepackage{microtype}      
\usepackage{lipsum}		
\usepackage{graphicx}
\usepackage{amssymb}
\usepackage{amstext}
\usepackage{amsmath}
\usepackage{amsthm}
\usepackage{subfig}  
\usepackage{stfloats}   
\usepackage{textcomp}  
\usepackage{cite}

\newcommand{\rvx}{\textbf{\emph{x}}}

\newcommand{\rvy}{\textbf{\emph{y}}}
\newcommand{\rvtheta}{\boldsymbol{\theta}}

\title{State-of-charge Estimation of a Li-ion Battery using Deep Learning and Stochastic Optimization}

\author{ \href{https://orcid.org/0000-0003-3557-1957}{\includegraphics[scale=0.06]{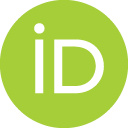}\hspace{1mm}Alexandre Barbosa de Lima} \\
	Polytechnic School of the University of S\~ao Paulo\\
	Department of Energy and Automation\\
	\texttt{alexandreblima@usp.br} 
	\And
	{\hspace{1mm}Maur\'icio B. C. Salles} \\
	Polytechnic School of the University of S\~ao Paulo\\
	Department of Energy and Automation  \\
	\texttt{mausalles@usp.br} \\
	\And
	{\hspace{1mm}Jos\'e Roberto Cardoso} \\
	Polytechnic School of the University of S\~ao Paulo\\
	Department of Energy and Automation  \\
	\texttt{jose.cardoso@usp.br} \\
}

\hypersetup{
pdftitle={A template for the arxiv style},
pdfsubject={q-bio.NC, q-bio.QM},
pdfauthor={David S.~Hippocampus, Elias D.~Striatum},
pdfkeywords={First keyword, Second keyword, More},
}

\begin{document}
\maketitle

\begin{abstract}
This article presents a novel empirical study for the estimation of the State of Charge (SOC) of a lithium-ion (Li-ion) battery which uses a deep learning model with three hidden layers. We model a series of ten vehicle drive cycles that were applied to a Panasonic 18650PF  Li-ion cell. Our results show that the choice of the optimization algorithm affects the model performance. The proposed model was able to achieve an error smaller than $1.0\%$ in all drive cycles.
\end{abstract}


\section{Introduction}\label{sec:intro}

In the last decade, government, industry and academia have given great importance to the electrification of the transport system, motivated by the need to reduce the emission of greenhouse gases. Hybrid electric vehicles, such as the Toyota Prius, or fully electric vehicles, such as the various Tesla models, the Nissan Leaf and the Chevy Bolt, are successful cases in the USA \cite{whit2012}.

The United Kingdom took a bold step in this direction in 2018. The Executive Summary of the document \textit{The Road to Zero: Next steps towards cleaner road transport and delivering our Industrial Strategy} (United Kingdom)  states that \cite{uk2018}:

\begin{quote}
	
	``Our strategy is built around a core mission: to put the UK at the forefront of the design and manufacturing of \textit{zero emission vehicles} and for all new cars and vans to be effectively zero emission by 2040. 
	
	(...)
	
	as we move to the mass adoption of ultra low emission vehicles, more infrastructure will be needed and we want to see improvements to the consumer experience of using it. Our vision is for current and prospective \textit{electric vehicle} drivers to be able to easily locate and access charging infrastructure that is affordable, efficient and reliable.
	
	(...)
	
	The electricity system of 2050 will look very different from today’s. There will be more low carbon generation, and \textit{new technologies such as battery storage} and onsite generation will play a bigger role. Decarbonising how we heat our homes and businesses will also bring further changes to the demands placed on the energy system.'' (our highlight)
	
\end{quote}	

The advancement of Electrical Energy Storage (EES) technologies enabled the emergence of the iPod, smartphones and tablets with lithium-ion (li-ion) batteries. Also, EES (cited above as ``battery storage'') will be one of the critical components of the new electricity grid, given the intermittent nature of renewable energy sources \cite{whit2012}, \cite{luo2015}. EES systems are necessary even when renewable sources are connected to the grid, because it is necessary to smooth the energy supply. For example, the EES of a building or factory can be charged during hours of reduced demand and supply/supplement energy demand during peak hours.


EES technology consists of the process of converting a form of energy (almost always electrical) to a form of storable energy, which can be converted into electrical energy when necessary. EES has the following functions: to assist in meeting the maximum electrical load demands, to provide time-varying energy management, to relieve the intermittency of renewable energy generation, to improve energy quality/reliability, to serve remote loads and vehicles, to support the realization of smart grids, improve the management of distributed/standby power generation and reduce the import of electricity during peak demand periods \cite{luo2015}, \cite{byrne2018}.

The efficient use of the Li-ion battery requires the supervision of a Battery Management System (BMS), as it is necessary that the battery operates under appropriate conditions of temperature and charge (State-Of-Charge (SOC)) \cite{huria2012}. The SOC can be measured using the Coulomb counting method of (\ref{eq:SOC})
\begin{equation}
	\label{eq:SOC}
	\text{SOC}  = \text{SOC}_0 - \frac{\int_{o}^{t} I_{\text{bat}}\,dt}{Q_n}
\end{equation} 
where $\text{SOC}_0$ is the initial value of SOC, $I_{\text{bat}}$ is the battery current and $Q_n$ is the nominal capacity in Ah. It should be noted that the cell temperature produces deleterious effects on the open circuit voltage, internal resistance and available capacity and can also lead to a rapid degradation of the battery if it operates above a given temperature threshold. Therefore, the modeling of the battery is of paramount importance, since it will be used by the BMS to manage the operation of the battery \cite{motapon2017}.

The recent literature suggests that the machine learning approach, based on deep learning algorithms  is the state of the art in this area \cite{huang2019}, \cite{zahid2018}, \cite{ren2018}, \cite{zhao2017}, \cite{chemali2018}, \cite{lstm-li-ion-2018}, \cite{Kollmeyer2017}, \cite{hannan2020}. 

One of the great challenges in deep learning is the optimization of the neural network. Although the Stochastic Gradient Descent (SGD) algorithm (and its variants) is very popular, a learning rate too small leads to painfully slow convergence, while a learning rate too high can destabilize the algorithm, causing oscillations or divergence \cite{goodfellow2016}. 

On the other hand, we have at our disposal algorithms with adaptive learning rates such as RMSProp \cite{tieleman2012} and Adamax \cite{goodfellow2016}, \cite{kingma2015}. 

To the best of our knowledge, there is no study on SOC estimation of Li-ion batteries that investigates the effect of choosing the optimization algorithm on the performance of the deep learning model. That is our contribution. The question we pose is: which optimization  algorithm should we choose? 

For example, \cite{huang2019}  ``postulates'' the use of the Adagrad algorithm \cite{duchi2011} for the optimization of a convolutional gated recurrent unit (CNN - GRU) architecture. However, as we will see in section \ref{sec:results}, sometimes the ``good old fashioned'' SGD can beat an algorithm with adaptive learning rules like Adagrad, at the cost of speed, as ``there is no free lunch''. Things are more complicated in the field of neural network optimization \cite{goodfellow2016}.

This work uses a deep learning model with three hidden layers. We model a series of ten vehicle drive cycles that were applied to a Panasonic 18650PF  Li-ion battery \cite{Kollmeyer2018}. More specifically, we compare the performance of the following algorithms: RMSProp \cite{tieleman2012}, Adamax \cite{kingma2015}, and SGD \cite{goodfellow2016}. Our results show that the choice of the optimization algorithm affects the model performance.
The issue of regularization (parameter norm penalties, early stopping, dropout, etc.) is outside the scope of this work. Nevertheless, the power of generalization of the proposed deep learning model  is high, since it was obtained a Mean Absolute Error (MAE)/Mean Squared Error (MSE) lower than $1.0\%$ for all of the ten drive cycles. 

The remainder of the paper is organized as follows.  Section \ref{sec:state} presents the state of the art and trends in Li-ion battery parameter estimation. Section \ref{sec:results} presents our experimental results. Finally, section \ref{sec:conclusions} presents our conclusions.

\section{State of the Art and Trends in Li-ion Battery Estimation}\label{sec:state}

Energy storage acts as a mediator between variable loads and variable sources. 

Hannan et al. \cite{hannan2016} present a detailed taxonomy of the types of energy storage systems taking into account the form of energy storage and construction materials: mechanical, electrochemical (rechargeable and flow batteries), chemical, electrical (ultracapacitor or superconducting magnetic coil), thermal and hybrid. Li-ion battery technology has attracted the attention of industry and academia for the past decade. This is mainly due to the fact that Li-ion batteries offer more energy, higher power density, higher efficiency and lower self-discharge rate than other battery technologies such as NiCd, NiMH, etc. \cite{motapon2017}.

There are two methods of battery modeling: i) model-driven and ii) data-driven (based on data that is collected from the device) \cite{ren2018}.

Electrothermal models, which belong to the category of model-driven methods, are commonly classified as: i) electrochemical or ii) based on Equivalent Circuit Models (ECM) \cite{huria2012}, \cite{motapon2017}.

Electrochemical models are based on partial differential equations \cite{jeon2014} and are able to represent thermal effects more accurately than ECM \cite{li2014}. However, the first class of models requires detailed knowledge of proprietary parameters of the battery manufacturer: cell area, electrode porosity, material density, electrolyte characteristics, thermal conductivity, etc. This difficulty can be eliminated by characterizing the battery using a thermal camera and thermocouples. But this solution is expensive, time consuming and introduces other challenges such as the implementation of dry air purge systems, ventilation, security, air and water supply, etc. Electrochemical models demand the use of intensive computing systems \cite{huria2012}.

On the other hand, the ECM-based approach has been used for computational/numerical analysis of batteries \cite{huria2012}. In this case, the objective is to develop an electrical model that represents the electrochemical phenomenon existing in the cell. The level of complexity of the model is the result of a compromise between precision and computational effort. Note that an extremely complex and accurate ECM may be unsuitable for application in embedded systems.


Chemali et al \cite{chemali2018} compared the performance of Deep Neural Networks (DNN) with those of other relevant algorithms that have been proposed since the second half of the 2000s. The article shows that the SOC estimation error obtained with deeep learning  is less than the following methods:

\begin{itemize}
	\item Model Adaptive-Improved Extended Kalman Filter (EKF) \cite{sepasi2014};
	\item Adaptive EKF with Neural Networks \cite{chark2010};
	\item Adaptive Unscented Kalman Filter (AUKF) with Extreme Machine Learning \cite{du2014};
	\item Fuzzy NN with Genetic Algorithm \cite{lee2014}; and
	\item Radial Bias Function Neural Network \cite{chang2013}.   
\end{itemize}

Estimating the SOC of lithium ion cells in a BMS by means of deep learning offers at least two significant advantages over model driven approaches, namely: i) neural networks are able to estimate the non linear functional dependence that exists between voltage, current and temperature (observable quantities) and unobservable quantities, such as SOC, with great precision  and ii) the problem of identifying ECM parameters is avoided.

\begin{figure}[htp]
	\centering
	\subfloat[]{
		\label{fig:temperature-SOC}%
		\includegraphics*[scale=0.3]{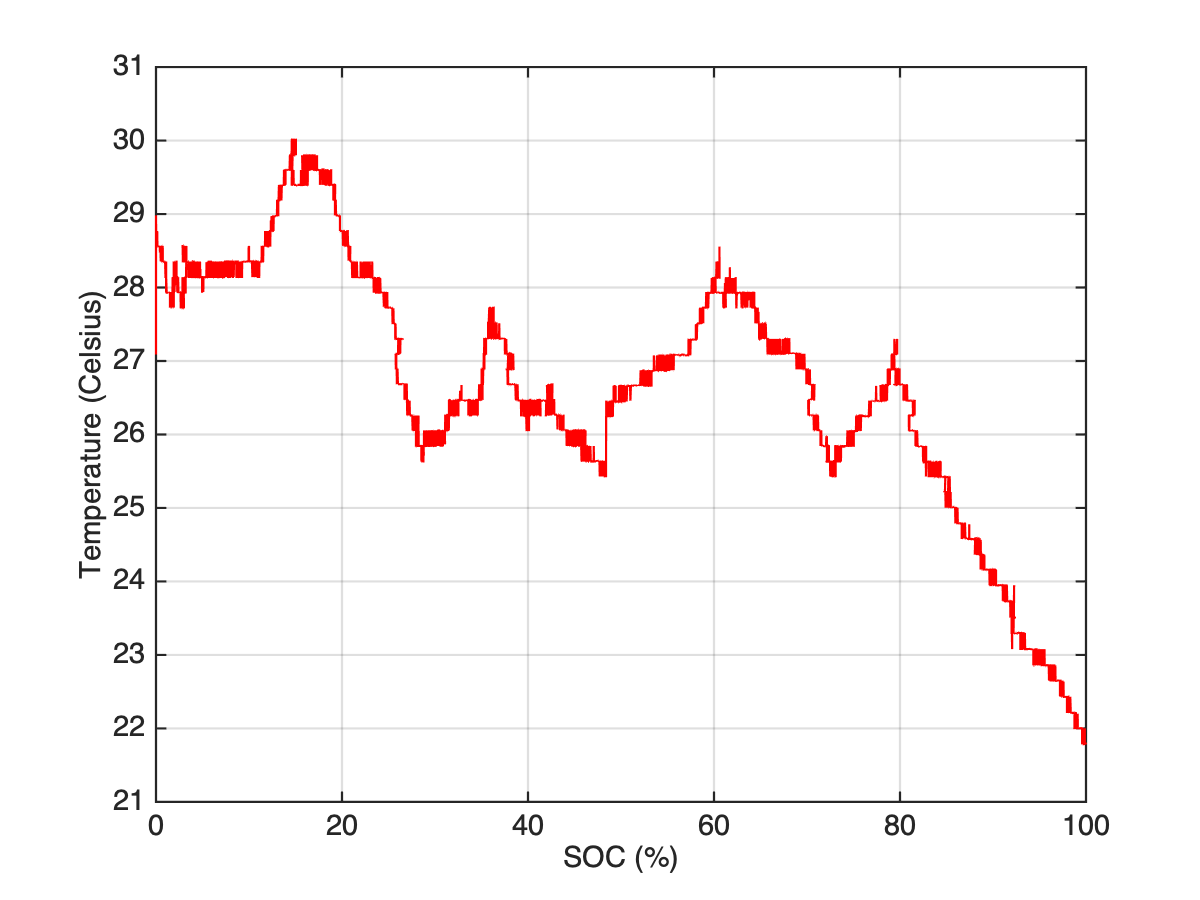}}		
	\subfloat[]{%
		\label{fig:Amphours-time}%
		\includegraphics*[scale=0.3]{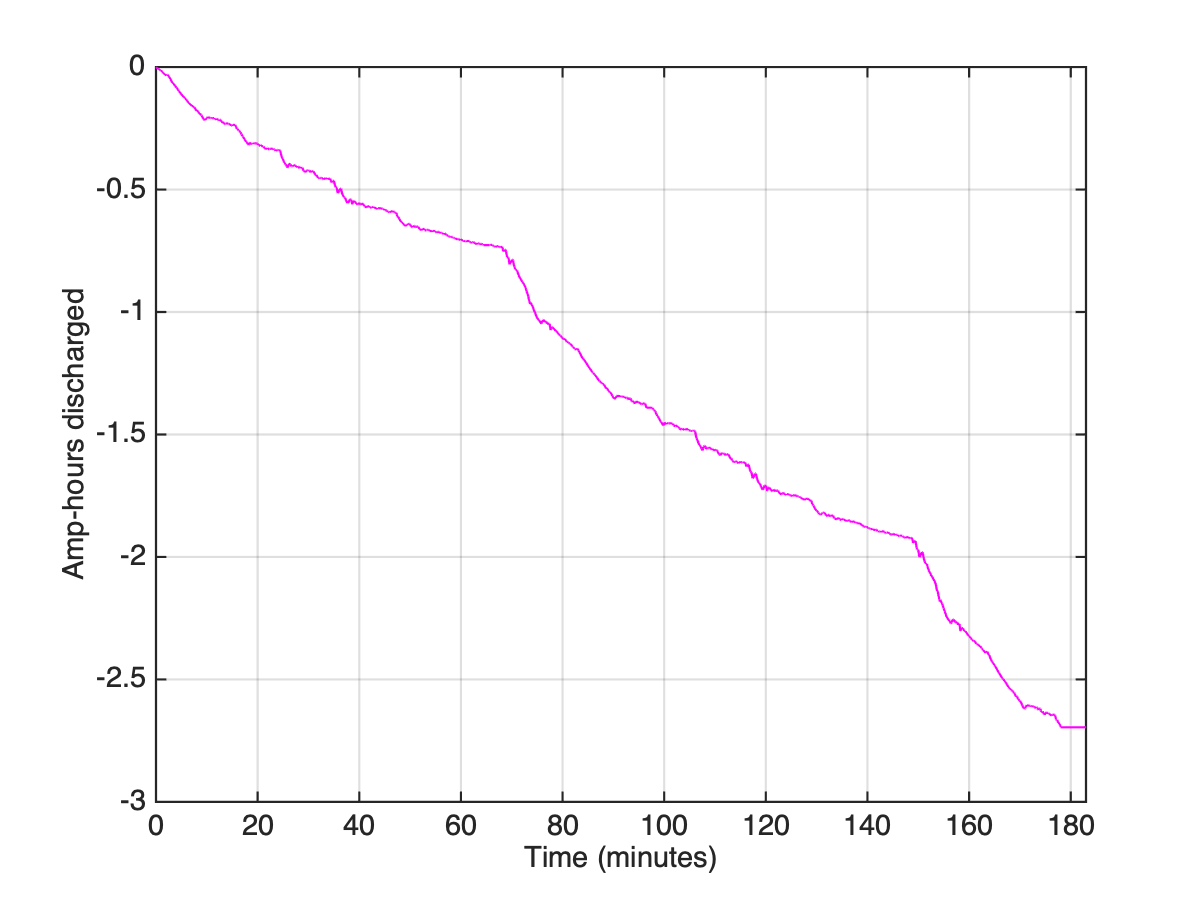}}\\
	\subfloat[]{%
		\label{fig:Voltage-time}%
		\includegraphics*[scale=0.3]{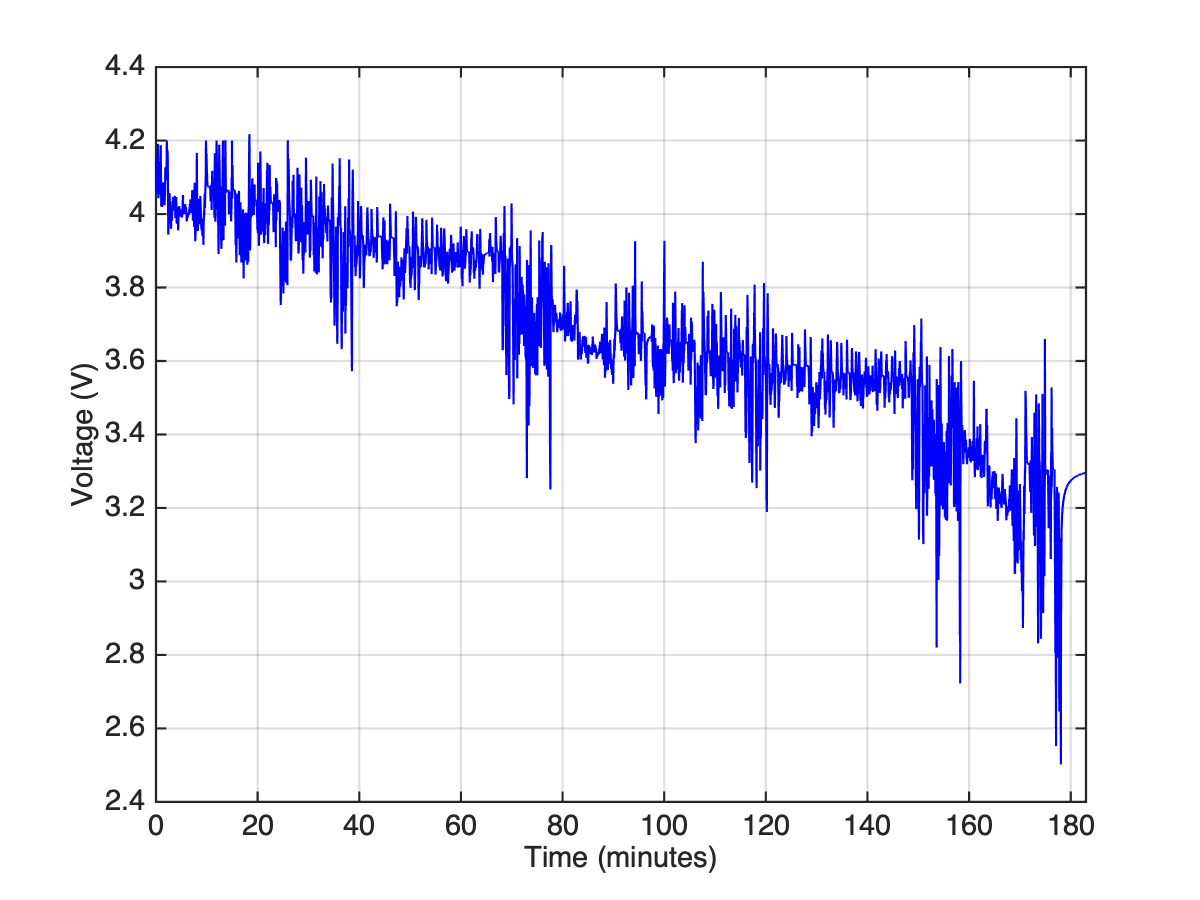}}
	\subfloat[]{%
		\label{fig:current-time}%
		\includegraphics*[scale=0.3]{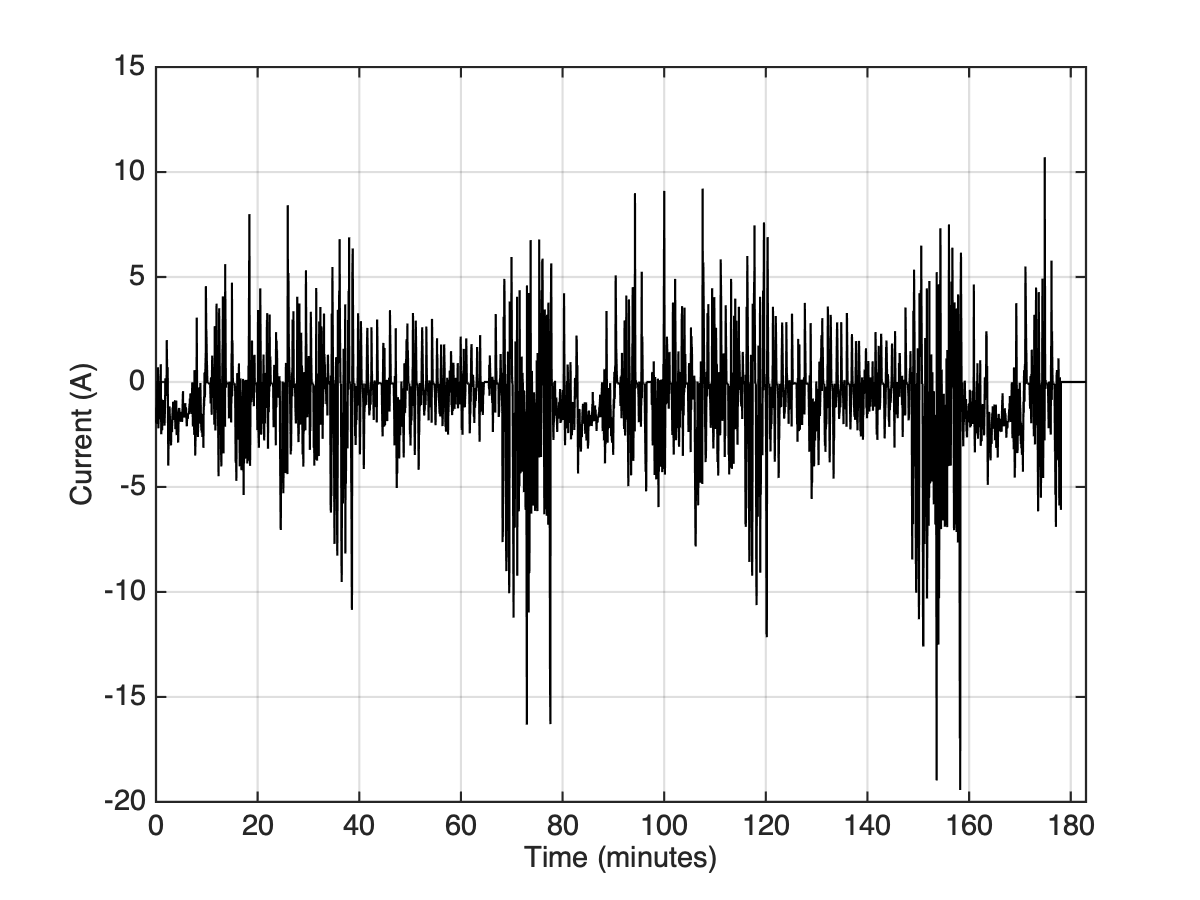}}\\
	\subfloat[]{%
		\label{fig:temperature-time}%
		\includegraphics*[scale=0.3]{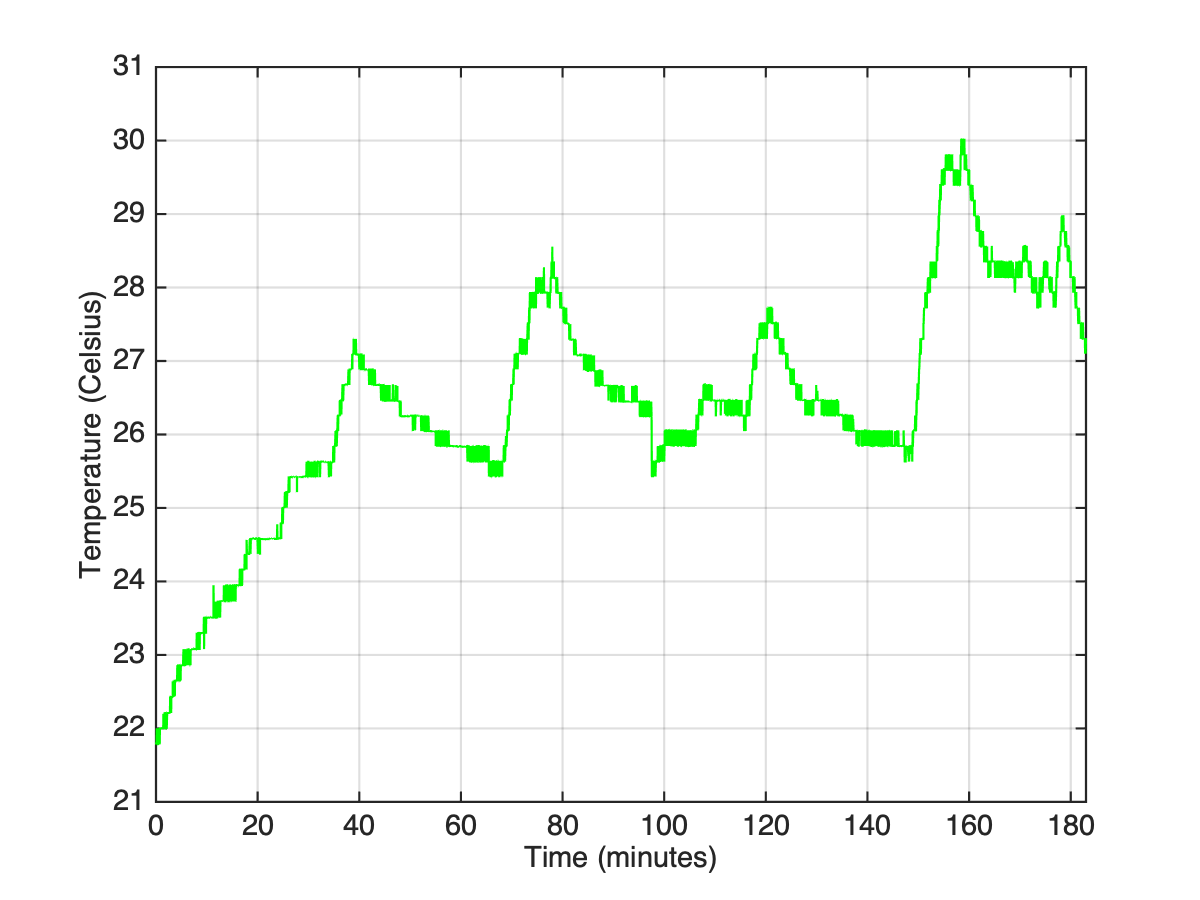}}
	\subfloat[]{%
		\label{fig:voltage-SOC}%
		\includegraphics*[scale=0.3]{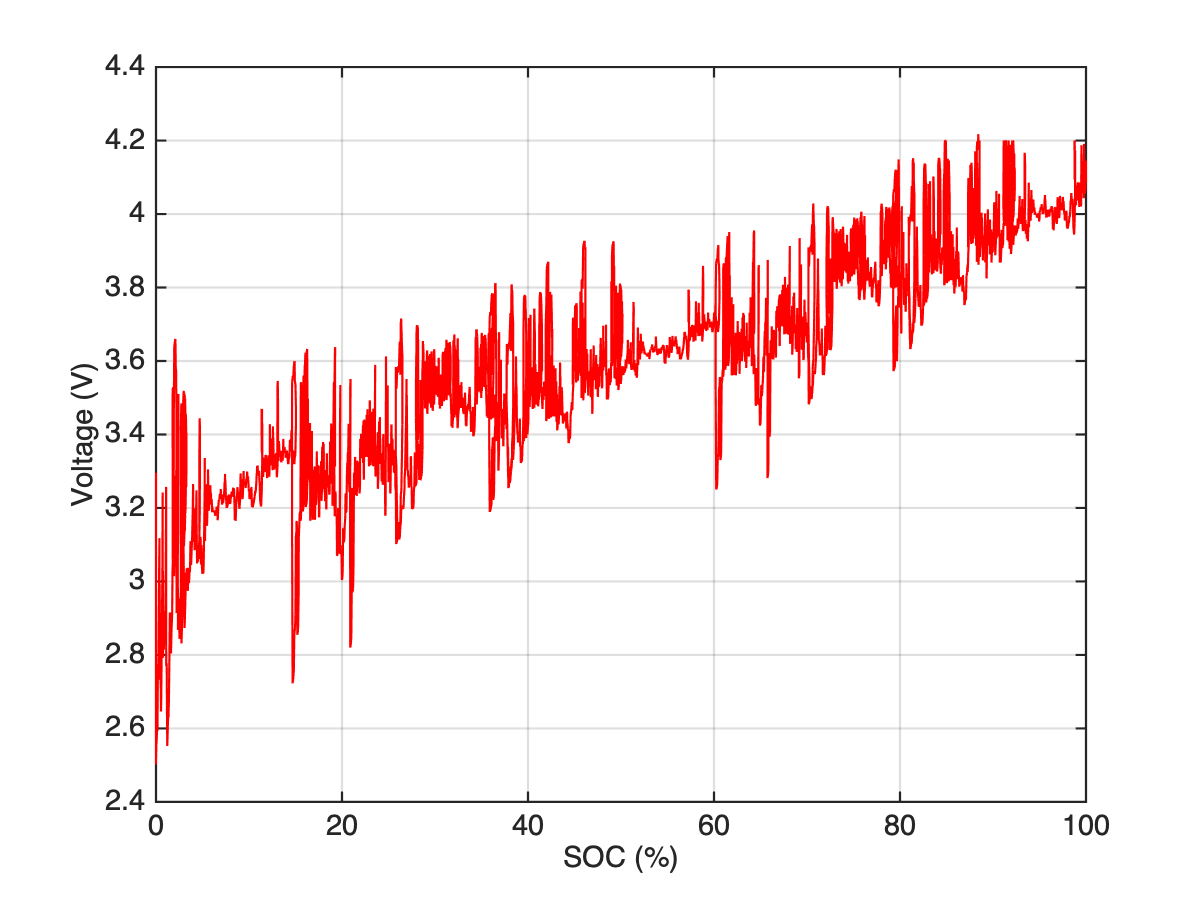}}\\
	\caption[]{
		in \subref{fig:temperature-SOC}, \subref{fig:Amphours-time}, \subref{fig:Voltage-time}, \subref{fig:current-time},
		\subref{fig:temperature-time}, and  \subref{fig:voltage-SOC} we have (for the drive Cycle 1): temperature ($^o$ C) vs SOC (\%), amp-hours discharged vs time (minutes), voltage (V) vs time (minutes), current (A) vs time (minutes), temperature ($^o$ C) vs time (minutes), and voltage (V) vs SOC (\%), respectively.}%
	\label{fig:curvas-caracteristicas}%
\end{figure}

\section{Experimental Results}\label{sec:results}

We selected the \textit{Panasonic 18650PF Li-ion Battery Data} of \cite{Kollmeyer2018}. 
This dataset contains a series of ten drive cycles: Cycle 1, Cycle 2, Cycle 3, Cycle 4, US06, HWFTa, HWFTb, UDDS, LA92, and Neural Network (NN).  Cycles 1-4 consist of a random mix of US06, HWFET, UDDS, LA92, and Neural Network drive cycles.  The drive cycle power profile is calculated from measurement for an electric Ford F150 truck with a 35kWh battery pack scaled for a single 18650PF cell.  We consider only the tests at the temperature of $25^o$ C, as the objective of this paper is not to assess the performance of the deep learning model at different temperatures, but to empirically investigate the question of stochastic optimization in the context of the problem of SOC estimation with deep neural nets.

For instance, Fig. \ref{fig:curvas-caracteristicas} shows the following 2.9 Ah Panasonic 18650PF Li-ion cell characteristic curves for Cycle 1.

\begin{itemize}
	\item temperature ($^o$ C) vs SOC (\%);
	\item amp-hours discharged vs time (minutes);
	\item voltage (V) vs time (minutes);
	\item current (A) vs time (minutes);
	\item temperature ($^o$ C) vs time (minutes); and
	\item voltage (V) vs SOC (\%).
\end{itemize}

We applied feature normalization on the input data using the formula

\begin{equation}
	\label{eq:feat-norm}
	\rvx_{\text{normalized}}  = \frac{\rvx - \mu_{\rvx}}{\sigma_{\rvx}} 
\end{equation} 

where $\mu_{\rvx}$ and $\sigma_{\rvx}$ denote the mean and standard deviation of $\rvx $.

\begin{figure}[h]
	\begin{center}
		\includegraphics*[scale = 0.30]{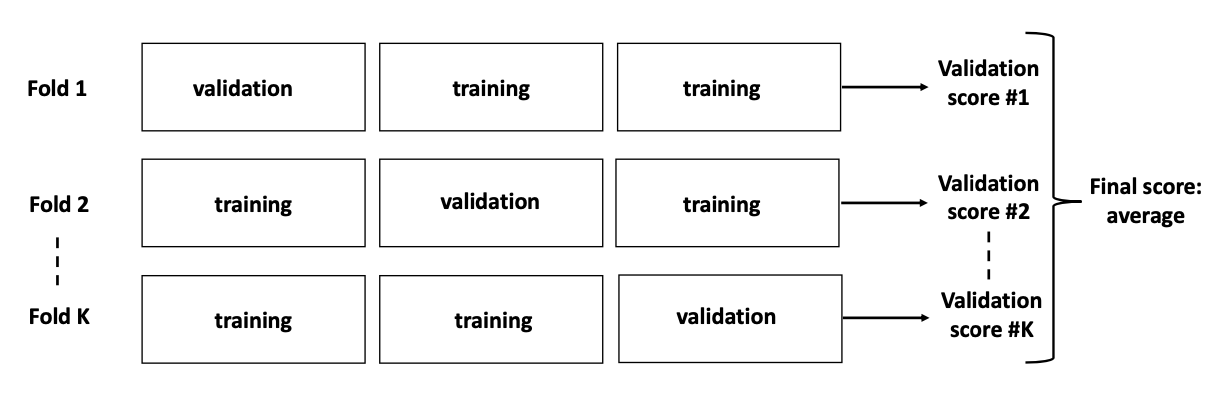}
	\end{center}	
	\caption{K-fold cross-validation.}
	\label{fig:kfold}
\end{figure}

We divided the datasets into training and validations sets. The validation error is estimated by taking the average validation error across $K=4$ trials. 
We use a simple, but popular solution, called $K$-fold cross-validaton (Fig. \ref{fig:kfold}), which consists of splitting the available training data into two partitions (training and validation), instantiating $K$ identical models, for each fold $k\in \{1,2,\ldots,K\}$, and training each one on the training partitions, while evaluating on the validation partition. The validation score for the model used is then the average of the $K$ validation scores obtained. This procedure allows network hyperparameters to be adjusted so that overfitting is mitigated \cite{goodfellow2016} . It is usual to use about $80\%$ of the data for the training set, and $20\%$ for the validation set. Note that the validation scores may have a high variance with regard to the validation split. Therefore, $K$-fold cross-validaton help us improve the reliability when evaluating the generalization power of the model. 

We used MAE and MSE as performance metrics for the generalization (test) error \cite{chemali2018}, \cite{goodfellow2016}. After the validation phase using $K$-fold cross-validaton, the deep learning model is trained using the entire training data and its performance is evaluated against an unseen test set. This the test phase (final phase).

The code was developed in \texttt{Python}, using  TensorFlow 2 and the Keras API \cite{tf}, \cite{keras}.

\begin{figure}[h]
	\begin{center}
		\includegraphics*[scale = 0.30]{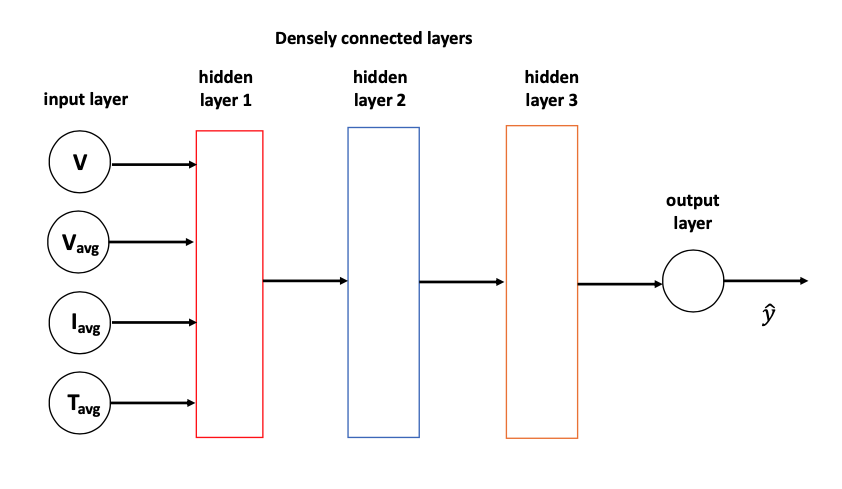}
	\end{center}	
	\caption{DFN with three (3) hidden layers. The output is denoted by $\hat{y}$. The network has a total of $133,121$ trainable parameters. The first, second, and third hidden layers have $1,280$, $65,792$, and $65,792$ parameters, respectively. The output layer has $257$ parameters (bias plus $256$ units of the last hidden layer).}
	\label{fig:dfn}
\end{figure}

Our simulations confirm the design matrix chosen by \cite{chemali2018}, as our best results were found with the following features (inputs): $x_1=V(t)$ (voltage in V), $x_2=\bar{V}(t)$ (average voltage in V), $x_3=\bar{I}(t)$ (average current in A), and $x_4=\bar{T}(t)$ (average temperature  in $^o$C), where $t$ denotes time in seconds. We used moving averages over $400$ past samples in order to smooth the last three features. This procedure improved the estimation of the SOC. The Deep Forward Network (DFN) has three hidden layers, each hidden layer having $256$ units (see Fig.\ref{fig:dfn}). We used the REctified Linear Unit (Relu) activation function, given by

\begin{equation}
	\label{eq:relu}
	g(z)  = \text{max}\{0,z\}.
\end{equation} 

Table \ref{tab:performance-optimizers-mae-mse} shows the SOC estimates obtained with SGD, RMSProp and Adamax. 
We used the default parameters for the Keras \texttt{SGD}, \texttt{RMSProp}, and \texttt{Adamax} classes. The \texttt{SGD}class, as well as the others, uses mini-batch gradient descent on $n$ training examples,
\begin{equation}
	\label{eq:mini-batch}
	\rvtheta_k = \rvtheta_{k-1} - \eta .\nabla J(     \theta;    \rvx^{(i:i+n)};  \rvy^{(i:i+n)}       )
\end{equation} 
where $\rvtheta$ denotes the network parameters, $\eta$ is the learning rate (also called step size), $\nabla(.)$ is the gradient, and $J(.)$ is the cost function. 

RMSProp is given by the following equations:
\begin{equation}
	\label{eq1:RMSPprop}
	\rvtheta_k = \rvtheta_{k-1} - \frac{\eta}{\sqrt{E[g^2]_k + \epsilon}  } . g_k
\end{equation} 
where
\begin{equation}
	\label{eq2:RMSPprop}
	E[g^2]_k = 0.9E[g^2]_{k-1} + 0.1 g_k^2
\end{equation} 
and 
\begin{equation}
	\label{eq3:RMSPprop}
	g_k = \nabla_{\rvtheta_k}J(\rvtheta_k).
\end{equation} 

Adamax is a variant of the ADAptive Moment Estimation (Adam) algorithm based on the infinity norm. Adam has the following update rule:
\begin{equation}
	\label{eq1:Adam}
	\rvtheta_k = \rvtheta_{k-1} - \frac{\eta}{\sqrt{\hat{v}_k} + \epsilon}  \hat{m}_k
\end{equation} 
where
\begin{equation}
	\label{eq2:Adam}
	\hat{m}_k = \frac{m_k}{1 - \beta_1^k}
\end{equation} 
and
\begin{equation}
	\label{eq3:Adam}
	\hat{v}_k = \frac{v_k}{1 - \beta_w^k}.
\end{equation}  
The variables $m_k$ and $v_k$ are estimates of the first and the second moments of the gradients, respectively:
\begin{equation}
	\label{eq4:Adam}
	m_k = \beta_1 m_{k-1} + (1 - \beta_1)g_k
\end{equation}
\begin{equation}
	\label{eq5:Adam}
	v_k = \beta_2 v_{k-1} + (1 - \beta_2)g_k^2.
\end{equation}    

The proponents of Adam indicate the default values of $0.9$ for $\beta_1$, $0.999$ for $\beta_2$, and $10^{-7}$ for $\epsilon$ \cite{keras}.

\begin{table}[h]
	\begin{center}
		\large
		\begin{tabular}{|c|c|c|c|c|c|c|}\hline\hline 
			\multicolumn{1}{|c|}{}	& \multicolumn{2}{c|} {\textbf{SGD}} & \multicolumn{2}{c}{\textbf{RMSProp}} &  \multicolumn{2}{|c|} {\textbf{Adamax}}                      \\ \hline
			\textbf{Drive Cycle}    &                   \textbf{MAE}       &              \textbf{MSE}            & \textbf{MAE}                 &  \textbf{MSE}                    &         \textbf{MAE}                                       & \textbf{MSE}                                        \\ \hline
			Cycle 1                              &       $0.81$                                 &   $1.04$                                  &  $0.96$                              &    $1.55$                               &      $0.45$              &    $0.43$         \\ \hline
			Cycle 2                             &        $0.65$                                 &  $0.77$                                   & $2.83$	                             &   $1.04$                              &     $0.47$                &    $0.47$              \\ \hline
			Cycle 3                             &       $0.68$                                  & $0.91$		                           &  $1.19$                                 & $2.24$		                         &     $0.57$              &    $0.63$         \\ \hline
			Cycle 4                             &      $0.48$                                   & $0.54$                                   &   $0.88$		                        & $1.29$	                            &     $0.54$            &  $0.61$          \\ \hline
			US06                                 &      $1.01$                                 & $2.08$       						       &   $1.11$		                         & $2.15$	                    &    $0.60$              &   $0.77$      \\ \hline
			HWFTa                             &       $0.40$                               & $0.35$                                     &   $0.39$	                             & $0.34$                        &      $0.41$               & $0.32$         \\ \hline
			HWFTb                             &       $0.41$                               & $0.36$                                     &   $1.33$	                             & $2.54$		                 &       $0.37$               & $0.28$        \\ \hline
			UDDS                                &       $0.68$                               & $0.70$                                    &  $1.13$		                           & $2.09$	                       &       $0.49$              & $0.44$       \\  \hline
			LA92                                & 		$0.59$		                         &	$0.70$	                                  & $0.83$	                                & $1.19 $		                &     	$0.63$	            & $0.71$		   \\ \hline
			NN                                    & 	  $1.08$	                            & $1.66$                                    & $0.68$	                              & $0.90$                        &        $0.48$              & $0.45$\\ \hline\hline
		\end{tabular}
	\end{center}
	\caption{MAE and MSE for all drive cycles. }
	\label{tab:performance-optimizers-mae-mse}
\end{table}

\begin{figure}[htp]
	\begin{center}
		\includegraphics*[scale = 0.50]{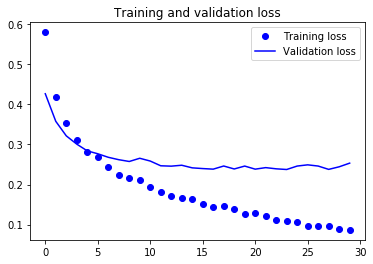}
	\end{center}
	\caption{Training and generalization errors behave differently. The horizontal axis represents the number of epochs. The vertical axis is the loss function. Note that the model starts to overfit around the fifth epoch.}
	\label{fig:training_validation_loss_cats_dogs_VGG16}
\end{figure}

\begin{figure}[!htp]
	\centering
	\subfloat[]{
		\label{fig:sgdmae}%
		\includegraphics*[scale=0.35]{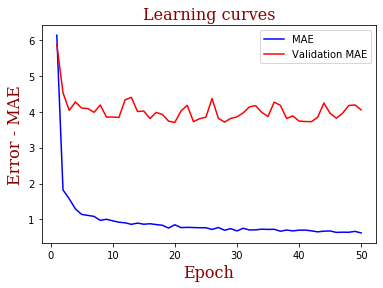}}		
	\subfloat[]{%
		\label{fig:rmspropmae}%
		\includegraphics*[scale=0.35]{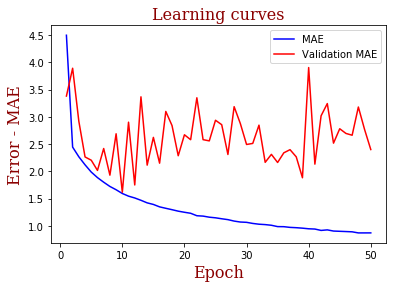}}
	\subfloat[]{%
		\label{fig:adamaxmae}%
		\includegraphics*[scale=0.35]{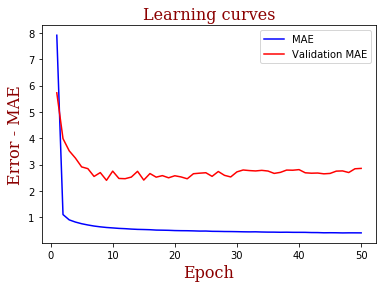}}
	\caption[]{
		in \subref{fig:sgdmae}, 
		\subref{fig:rmspropmae}, and
		\subref{fig:adamaxmae} we have: 
		MAE (SGD) vs Epoch, 
		MAE (RMSProp) vs Epoch, and
		MAE (Adamax) vs Epoch, 
		respectively.}%
	\label{fig:learn-curves}%
\end{figure}

\begin{figure}[!htp]
	\centering
	\subfloat[]{%
		\label{fig:sgdpred}%
		\includegraphics*[scale=0.35]{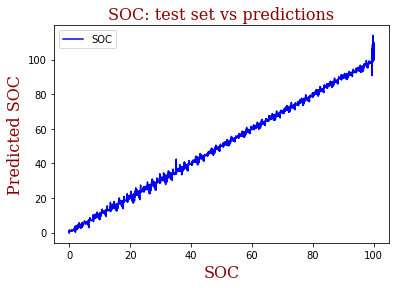}}
	\subfloat[]{%
		\label{fig:rmsproppred}%
		\includegraphics*[scale=0.35]{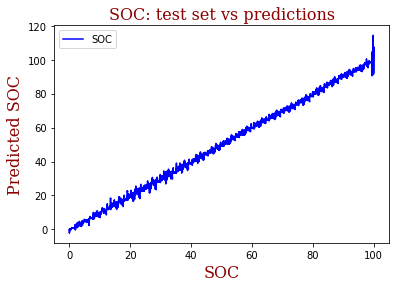}}
	\subfloat[]{%
		\label{fig:adamaxpred}%
		\includegraphics*[scale=0.35]{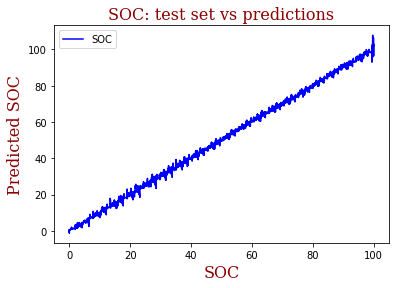}}\\
	\caption[]{
		in 
		\subref{fig:sgdpred}, 
		\subref{fig:rmsproppred}, and
		\subref{fig:adamaxpred} we have: 
		Predicted SOC (SGD) vs Measured SOC,  
		Predicted SOC (RMSProp) vs Measured SOC, 
		and Predicted SOC (Adamax) vs Measured SOC, respectively.}%
	\label{fig:pred-curves}%
\end{figure}

The results of Table \ref{tab:performance-optimizers-mae-mse} show that:
\begin{itemize}
	\item the choice of the optimization algorithm must be made on a case-by-case basis. Although the family of algorithms with adaptive learning rates (RMSProp, Adam, Adadelta, etc.) has become very popular in the deep learning community nowadays, one can not claim, a priori, that a given optimization algorithm is the best algorithm \cite[p. 302]{goodfellow2016};
	\item the choice of the optimization algorithm affects the model performance. Also note the good performance obtained with the hidden 3-layer DFN model, who was able to achieve a MAE/MSE smaller than $1.0\%$ in all drive cycles.
\end{itemize}


Figs. \ref{fig:learn-curves} and \ref{fig:pred-curves} show, for the NN drive cycle, the learning curves (validation phase) and the prediction curves (test phase) for the SOC, respectively.

\section{Conclusions}\label{sec:conclusions}

We present a preliminary empirical study on the impact of the optimization algorithm on the performance of the deep learning model, in the context of the estimation of the SOC of a Li-ion battery in ten distinct scenarios. For this, we used the dataset \cite{Kollmeyer2018}.

Our results indicate that: i) the choice of the optimization algorithm must be made on a case-by-case basis, and ii) the choice of the optimization algorithm affects the model performance.

Although the family of algorithms with adaptive learning rates (RMSProp, Adam, Adadelta, etc.) has become very popular in the deep learning community nowadays, one can not claim, a priori, that a given optimization algorithm is the best algorithm.

Finally, it is important to highlight the good performance obtained with the hidden 3-layer DFN model, who was able to achieve a MAE/MSE smaller than $1.0\%$ in all drive cycles.

%
%

\end{document}